\documentclass[pre,preprint,showpacs,preprintnumbers,amsmath,amssymb]{revtex4}

\usepackage[dvips]{graphicx}% Include figure files
\usepackage{dcolumn}% Align table columns on decimal point
\usepackage{bm}% bold math

\usepackage[english]{babel}

\usepackage[latin1]{inputenc}

\usepackage{times}
\usepackage[T1]{fontenc}
\begin{document}
\title{Continuum theory of tilted chiral smectic phases.}

\author{J. P. MARCEROU }
\email{marcerou@crpp-bordeaux.cnrs.fr}
\affiliation{ Centre de Recherches Paul Pascal, 115, Av.
Albert-Schweitzer, 33600 Pessac, France}

\homepage{http://www.crpp-bordeaux.cnrs.fr}

\date{\today}
\begin{abstract}

We demonstrate that the sequence of distorted commensurate phases observed in tilted chiral smectics is explained by the gain in electrostatic energy due to the lock-in of the unit cell to a number of layers which is the integer closest to the ratio pitch over thickness of the subjacent Sm-C$^*_\alpha$ phase. We also explain the sign change of the helicity in the middle of the sequence by a balance between two twist sources one intrinsic and another due to the distortion of the Sm-C$^*_\alpha$.

\end{abstract}

\pacs{61.30.Dk \quad 64.60.Ej \quad 64.70.mf}
\maketitle
\section{Introduction}

\begin{figure}[!h]
  \includegraphics[width=6cm]{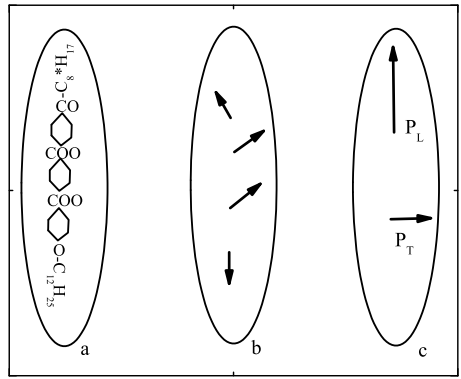}\\
  \caption{dipolar distribution from real molecule to the Boulder model.}\label{Clark}
\end{figure}

Chiral smectics have two proper characteristics, they are allowed to be ferroelectric and to present a helical precession of the optical axes around the layer normal when a tilt of the molecules appears in the layers \cite{Meyer}. Like other mesophases \cite{PGDG}, they are ferroquadrupolar phases in the sense that a large amount of the individual dipoles orient themselves collectively in the bulk and sum up in an antiparallel way to give sizeable effects like the flexoelectricity \cite{flexo}.
The liquid crystals molecules bear polar links like C=O, N-O, C$\equiv$N and delocalized electrons (figure \ref{Clark}a), so they present a distribution of dipoles all along their skeleton (figure \ref{Clark}b). The Boulder group has shown that the molecular dipoles can be approximated without loss of generality by a longitudinal one $\vec{P_L}$ and a transverse one $\vec{P_T}$   \cite{PClark} with amplitudes of several debyes.

Most of the literature in this field has dealt only with the transverse polarization which is at the origin of the ferroelectricity \cite{Meyer,japs,11,Fucku12,13,Z16,O17,18,19,20,21,L22,beta,Martinot}, and only a few have recognized the importance of the longitudinal one \cite{PClark,Fucku14}.

In this paper, we will try to explain the mechanisms at the origin of the formation of the different tilted chiral smectics.
\section{The chiral smectic phases}
By order of increasing complexity one encounters the following phases which structure is best described by the distorted clock model mainly developed from the data of resonant X-rays scattering experiments \cite{Mach,3,4,5,HT1,CC,Gleeson}.
\subsection{Sm-A}
\begin{figure}[!h]
  \includegraphics[width=6cm]{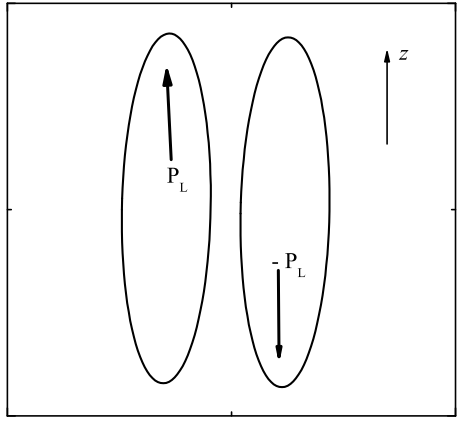}\\
  \caption{Schematic arrangement of molecules in the Sm-A phase. The transverse dipoles vanish while the longitudinal ones are in equal numbers up or down. Note that each of the sketched molecules represent symbolically one half of all molecules in the bulk, it is not a microscopic view.}\label{figSmA}
\end{figure}

The initial phase which precedes the various tilted phases at higher temperature is the smectic A (Sm-A). The molecules are normal to the layers. The transverse dipoles average to zero due to a uniform rotation about the long axis. The longitudinal dipoles adopt equiprobable up and down orientations (fig. \ref{figSmA}) ensuring that there is no macroscopic polarization but a macroscopic uniaxial quadrupole $\Theta_{ij}$. The uniaxial orientational order parameter (OOP) is expressed as $S_{ij}   =  n_i  n_j - \frac{1}{3}\delta _{ij}$ where $\overrightarrow{n}$ is the director. When it is written in a frame for which the normal to the smectic layers is taken as the $z$ direction it reads  \cite{PGDG}~:

\begin{equation}
 S_{ij}   = \left( {{\begin{array}{c c c}
 -1/3  & 0  & 0  \\
 0  & -1/3  & 0  \\
 0  & 0 & +2/3
\end{array} }} \right), \qquad \Theta_{ij} = \Theta_a S_{ij}
\end{equation}

\subsection{Sm-C$^*$}
\begin{figure}[!h]
  \includegraphics[width=8cm]{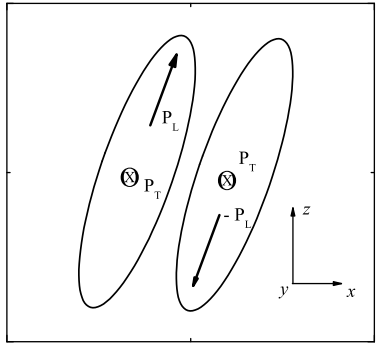}\\
  \caption{Tilted molecules in a Sm-C$^*$ layer, with the longitudinal dipoles in equal number in opposite directions while the hindered rotation leaves an average transverse dipole.}\label{figSmC}
\end{figure}
If the preferred layer thickness decreases with temperature and becomes smaller than the length of the molecules, they have to tilt in one direction giving in the simplest case the phase predicted by Meyer \cite{Meyer}, the smectic C$^*$ (Sm-C$^*$) where all the molecules are parallel (figure \ref{figSmC}). The transverse dipoles give birth to the macroscopic polarization $P_S$ when summed up over at least ten layers. The longitudinal ones have to average to zero but they still sum up in a macroscopic quadrupole which main axis is tilted with respect to the layer normal. If one approximates the OOP of the Sm-C$^*$ as being the same $S_{ij}$ as in the Sm-A tilted at an angle $\theta$ with respect to the layer normal in the azimuthal direction $\Phi_0$, one gets~:
%\begin{widetext}
 \begin{eqnarray}\label{3M}
\nonumber  Q_{ij} &=& \left( 1 - \frac{3}{2}\,\sin^2\theta \right) \left( \begin{array}{c c c}
 -1/3  & 0  & 0  \\
 0  & -1/3  & 0  \\
 0  & 0 & +2/3
\end{array} \right) \\ \nonumber \\
&+& \frac{1}{2}\sin^2\theta
\left( \begin{array}{c c c}
 \cos 2\Phi_0  & \sin 2\Phi_0  & \ 0\   \\
 \sin 2\Phi_0  & -\cos 2\Phi_0  & \ 0\   \\
 0  & 0 &\  0\
\end{array} \right)\\ \nonumber
\\ \nonumber
 ~ &-& \sin \theta \cos \theta
 \left( \begin{array}{c c c}
  0 & 0  & \ \cos \Phi_0\   \\
 0  & 0  & \ \sin \Phi_0\   \\
 \cos \Phi_0  & \sin \Phi_0 &\  0\
\end{array} \right)
\end{eqnarray}
%\end{widetext}

Due to the chirality, the structure precesses around the layer normal $z$ following the law $\Phi_0 = q_1 z$ with a pitch in the micron range.

The macroscopic quadrupole $\Theta_{ij}$ will be to first order proportional to $Q_{ij}$. The more realistic case of biaxial $S_{ij}$ is treated in appendix B and keeps the same symmetry as in equ.(\ref{3M}) with slightly involved factorized coefficients.

This expression with three basic matrices will be found in all the tilted phases and is fundamental for the continuum theory we have developed.

\subsection{Sm-C$^*_A$}

\begin{figure}[!h]
  \includegraphics[width=6cm]{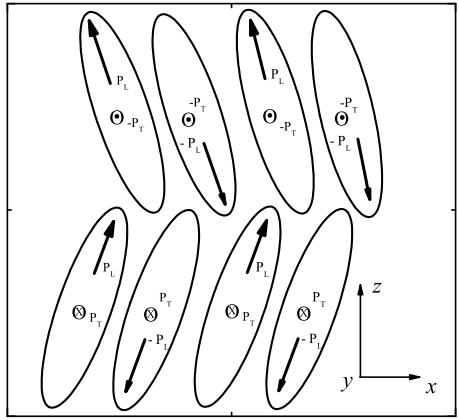}\\
  \caption{in the anticlinic Sm-C$^*_A$ phase, the transverse and longitudinal dipoles are compensated when pairing the layers, so the phase is ferro-quadrupolar.}\label{figSmCA}
\end{figure}

In the anticlinic phase with a period of two layers (figure \ref{figSmCA}), both longitudinal and transverse dipoles contribute to a macroscopic biaxial quadrupole which has the three C2  directions $x$, $y$ and $z$ as symmetry axes. This phase is misleadingly referred to be antiferroelectric due to the alternate orientations of $\vec{P_T}$, one should notice that there is also an alternance of $\vec{P_L}$ that leads to an other periodic array in the $x$ direction \cite{PClark,Fucku14}. So it is better to characterize this phase by its quadrupole where $\vec{P_T}$ contributes to $\Theta_{yy}$ and $\vec{P_L}$ to $\Theta_{xx}$ and $\Theta_{zz}$

As the Sm-C$^*_A$ is built by combining $\varphi = \Phi_0$ and $\varphi = \Phi_0 + \pi$, the OOP reads~:
%\begin{widetext}
 \begin{eqnarray}\label{2M}
  \nonumber Q_{ij} &=& \left( 1 - \frac{3}{2}\,\sin^2\theta \right) \left( \begin{array}{c c c}
 -1/3  & 0  & 0  \\
 0  & -1/3  & 0  \\
 0  & 0 & +2/3
\end{array} \right)  \\ \\ \nonumber &+& \frac{1}{2}\sin^2\theta
\left( \begin{array}{c c c}
 \cos 2\Phi_0  & \sin 2\Phi_0  & \ 0\   \\
 \sin 2\Phi_0  & -\cos 2\Phi_0  & \ 0\   \\
 0  & 0 &\  0\
\end{array} \right)
\end{eqnarray}
%\end{widetext}
with again a precession $\Phi_0 = q_2 z$ where $q_2$ has the opposite sign to $q_1$ in a given compound.
\subsection{Sm-C$^*_{Fi1}$}
\begin{figure}[!h]
  \includegraphics[width=6cm]{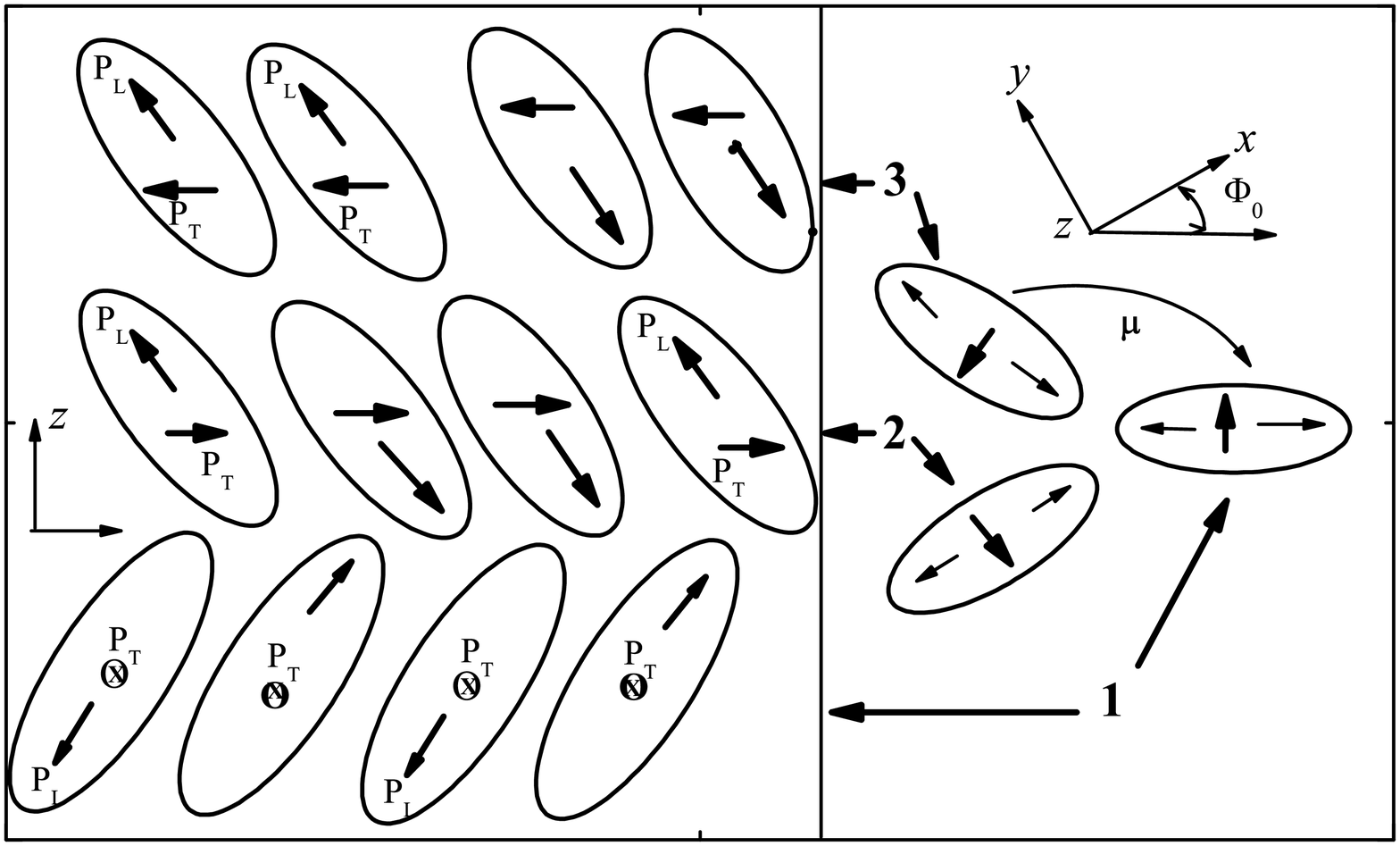}\\
  \caption{left : side view of the 3 layers unit cell of a Sm-C$^*_{Fi1}$. at right top view assuming a clockwise rotation, the difference in azimuthal angles between layers 1 and 2 or between 2 and 3 is taken as $\mu$ \cite{HT1,nous}. The in-plane projection of the director of layer 1 makes the angle $\Phi_0$ with x.}\label{figFi1}
\end{figure}
This phase presents a unit cell commensurate to three layers with unequal changes of the azimuthal angle from layer to layer ($\Delta\varphi = \mu$ or $2(\pi-\mu)$) see e.g figure (\ref{figFi1}) with the consequences that there is a neat polarization at larger scale (the Sm-C$^*_{Fi1}$ is ferrielectric) and a macroscopic precession of the structure around the layer normal($\Phi_0$ = q$^\prime_1 z$). All these informations can be gathered when writing the OOP of the phase~:
%\begin{widetext}
\begin{eqnarray}\label{OOP1}
\nonumber  Q_{ij} &=& \left( 1 - \frac{3}{2}\,\sin^2\theta \right) \left( \begin{array}{c c c}
 -1/3  & 0  & 0  \\
 0  & -1/3  & 0  \\
 0  & 0 & +2/3
\end{array} \right) \\ \nonumber \\ &+& \frac{J}{2}\sin^2\theta
\left( \begin{array}{c c c}
 \cos 2\Phi_0  & \sin 2\Phi_0  & \ 0\   \\
 \sin 2\Phi_0  & -\cos 2\Phi_0  & \ 0\   \\
 0  & 0 &\  0\
\end{array} \right)\\ \nonumber
\\ \nonumber
 ~ &-& I\;\sin \theta \cos \theta
 \left( \begin{array}{c c c}
  0 & 0  & \ \cos \Phi_0\   \\
 0  & 0  & \ \sin \Phi_0\   \\
 \cos \Phi_0  & \sin \Phi_0 &\  0\
\end{array} \right)
\end{eqnarray}
%\end{widetext}
taking the definitions of $\mu$ and $\Phi_0$ given in the figure (\ref{figFi1}), one finds \cite{nous} that the polarization $P_S$ is proportional to $I = (1 + 2 \cos \mu)/3$ while the macroscopic quadrupole $\Theta_{ij}$ is a function of $\theta$, $I$ and $J = (1 + 2 \cos 2 \mu)/3$ and is tilted with respect to the layer normal.
\subsection{Sm-C$^*_{Fi2}$}
\begin{figure}[!h]
  \includegraphics[width=6cm]{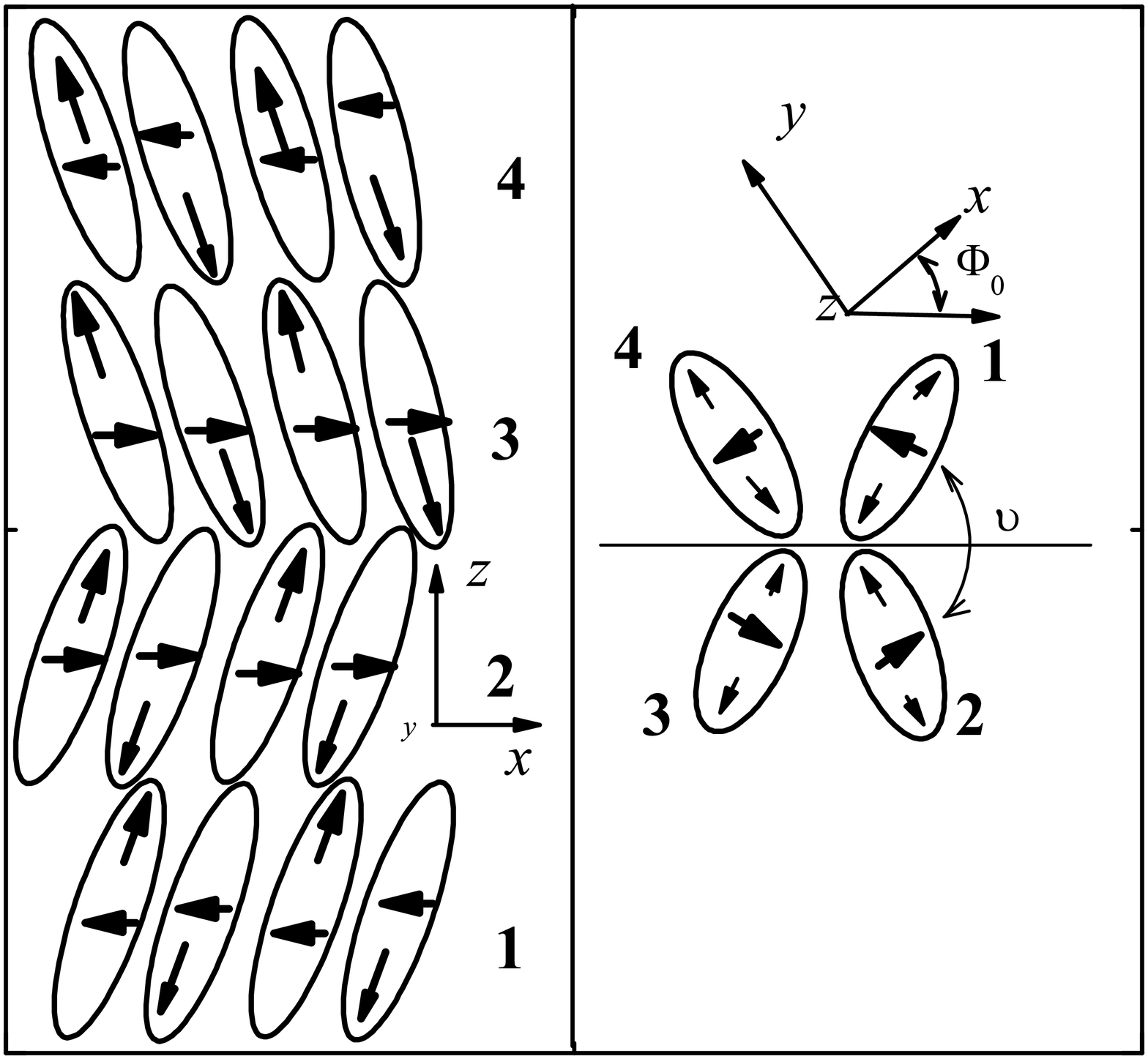}\\
  \caption{left : side view of the 4 layers unit cell of a Sm-C$^*_{Fi2}$. at right top view assuming a clockwise rotation, the difference in azimuthal angles between layers 1 and 2 or between 3 and 4 is taken as $\upsilon$ \cite{HT1,nous}. The in-plane projection of the bissectrix of layers 1 and 2 makes the angle $\Phi_0$ with x.}\label{figFi2}
\end{figure}
The unit cell is commensurate to four layers with unequal changes of the azimuthal angle from layer to layer ($\Delta\varphi = \upsilon$ or $\pi-\upsilon$) see e.g figure (\ref{figFi2}) with now no neat polarization at larger scale (the Sm-C$^*_{Fi2}$ is not ferrielectric !) and a macroscopic precession of the structure around the layer normal ($\Phi_0$ = q$^\prime_2 z$, in almost all the studied compounds, $q_1$ and q$^\prime_2$ have the same sign while $q_2$ and q$^\prime_1$ have the opposite). All these informations are gathered in the OOP of the phase~:
%\begin{widetext}
\begin{eqnarray}\label{OOP2}
 \nonumber Q_{ij} &=& \left( 1 - \frac{3}{2}\,\sin^2\theta \right) \left( \begin{array}{c c c}
 -1/3  & 0  & 0  \\
 0  & -1/3  & 0  \\
 0  & 0 & +2/3
\end{array} \right) \\ \\ \nonumber &+& \frac{J}{2}\sin^2\theta
\left( \begin{array}{c c c}
 \cos 2\Phi_0  & \sin 2\Phi_0  & \ 0\   \\
 \sin 2\Phi_0  & -\cos 2\Phi_0  & \ 0\   \\
 0  & 0 &\  0\
\end{array} \right)
\end{eqnarray}
%\end{widetext}
with the definitions of $\upsilon$ and $\Phi_0$ given in the figure (\ref{figFi2}), one finds \cite{nous} that the macroscopic quadrupole $\Theta_{ij}$ which has the layer normal as one of its eigenaxes is a function of $\theta$ and $J = - \cos \upsilon$.

\subsection{Sm-C$^*_{d6}$}
A last commensurate phase with six layers has been predicted by H\&T \cite{HT1,HT2} and recently evidenced by Shun Wan et al \cite{CC2010}. We will not develop on it but it has a symmetry close to that of Sm-C$^*_{Fi2}$ and similar properties.
\subsection{Sm-C$^*_\alpha$}
Last but not least, this phase shows a periodic precession with a short period which is not commensurate to the layer thickness. Its macroscopic OOP is simply uniaxial~:
\begin{equation}\label{OOPa}
 Q_{ij} = \left( 1 - \frac{3}{2}\,\sin^2\theta \right) \left( \begin{array}{c c c}
 -1/3  & 0  & 0  \\
 0  & -1/3  & 0  \\
 0  & 0 & +2/3
\end{array} \right)
\end{equation}
H\&T have shown that the Sm-C$^*_\alpha$ is fundamental for the obtention of the commensurate subphases. In short the incommensurate period varies continuously with the temperature taking values comprised in a subset of the interval of 2 to 8 layers. They have shown \cite{HT1,HT2} that when this period gets close to an integer number of layers, the system prefers to lock-in at this integer value at the expense of the twist energy compensated by some other gains. They have proposed that this gain scales at $J^2$ due to an anisotropy of in-plane elastic energy. What we propose here is that the lock-in allows the onset of macroscopic quadrupole and sometimes dipole with $J^2$ and $I^2$ contributions that explain the development of the full set of subphases.
\section{The sequence of tilted phase}

\subsection{The macroscopic and microscopic orientational order parameters (OOP \& oop)}
We have just seen that all the phases described by the distorted clock model are characterized by their macroscopic orientational order parameter (OOP) $Q_{ij}$ which general form valid in all phases has been given in equation (\ref{OOP1}). It is defined on a scale of at least ten layers like the quadrupole $\Theta_{ij}$ and the polarization $\overrightarrow{P_S}$. We have also recalled the fundamental statement of H\&T that the helicity of the Sm-C$^*_\alpha$ governs the appearance of other phases. So one has to develop the theory of the Sm-C$^*_\alpha$ phase and its transition from the Sm-A. For that we consider that each layer in a tilted smectic phase is such that the director makes an angle $\theta$ with $z$ while its in-plane projection makes the angle $\varphi$ with $x$ ; so we express the result of the rotation of $S_{ij}$ in the $xyz$ frame as the tensor $s_{ij}$ which is the microscopic orientational order parameter (oop) of the layer~:

%\begin{widetext}
 \begin{eqnarray}\label{inv}
\nonumber  s_{ij} &=& \left( 1 - \frac{3}{2}\,\sin^2\theta \right) \left( \begin{array}{c c c}
 -1/3  & 0  & 0  \\
 0  & -1/3  & 0  \\
 0  & 0 & +2/3
\end{array} \right) \\ \nonumber \\ &+& \frac{1}{2}\sin^2\theta
\left( \begin{array}{c c c}
 \cos 2\varphi  & \sin 2\varphi  & \ 0\   \\
 \sin 2\varphi  & -\cos 2\varphi  & \ 0\   \\
 0  & 0 &\  0\
\end{array} \right) \\ \nonumber
\\ \nonumber
 ~ &-& \sin \theta \cos \theta
 \left( \begin{array}{c c c}
  0 & 0  & \ \cos \varphi\   \\
 0  & 0  & \ \sin \varphi\   \\
 \cos \varphi  & \sin \varphi &\  0\
\end{array} \right)
\end{eqnarray}
%\end{widetext}

As already reported in \cite{JRL,LD} the oop  splits into three traceless invariants, namely  a bulk 3D-uniaxial tensor which depends only on $\theta$, and two in-plane tensors respectively 2D-biaxial and  2D-uniaxial depending also on $\varphi$. This local tensor will be used to compute the bulk OOP of each tilted phase by including the $z$ dependence of the azimuth angle $\varphi$.

\subsection{Landau - de Gennes free energy}

The free energy density describing the phase transition from the Sm-A to tilted phases can be written as a power series of the local  oop $s_{ij}$~:

\begin{widetext}
\begin{equation}\label{LdG}
    F_1 = \frac{1}{2}a_{ijkl}s_{ij}s_{kl} + \frac{1}{3}\Omega_{ijklmn}s_{ij}s_{kl}s_{mn} + \frac{1}{4}b_{ijklmnop}s_{ij}s_{kl}s_{mn}s_{op}
\end{equation}
\end{widetext}
Following the Smith and Rivlin theorem \cite{SmRv} we express the tensorial coefficients $a_{ijkl}$, $\Omega_{ijklmn}$ and $b_{ijklmnop}$ as products of the elementary tensors like the Kronecker $\delta_{ij}$, the Sm-A OOP $S_{ij}$, the vacuum tensor $V_{ij}$ and as we deal with chiral compounds the fully antisymmetric Levi-Civita odd tensor $e_{ijk}$.

After some tedious calculations \cite{JRL,LD} one gets rather simple results which are functions of the invariants introduced in (\ref{inv})~:
\begin{widetext}
\begin{equation}\label{n2}
    \frac{1}{2}a_{ijkl}s_{ij}s_{kl} = \frac{1}{2}a_1\;s_{zz}^2 + \frac{1}{2}a_2\;\left((s_{xx}-s_{yy})^2+4s_{xy}^2\right) + \frac{1}{2}a_3\;(s_{xz}^2+s_{yz}^2)
\end{equation}

\begin{eqnarray}\label{n3}
    \frac{1}{3}\Omega_{ijklmn}s_{ij}s_{kl}s_{mn} &=& \frac{1}{3}\Omega_1\;s_{zz}\;(s_{xz}^2+s_{yz}^2) + \frac{1}{3}\Omega_2\;s_{zz}\;\left((s_{xx}-s_{yy})^2+4s_{xy}^2\right) \nonumber \\ &+& \frac{1}{3}\Omega_3\;\left((s_{xx}-s_{yy})\;(s_{xz}^2-s_{yz}^2) - 4 s_{xy} s_{xz} s_{yz}\right)
\end{eqnarray}

\begin{eqnarray}\label{n4}
    \frac{1}{4}b_{ijklmnop}s_{ij}s_{kl}s_{mn}s_{op} &=& \frac{1}{4}b_1\;s_{zz}^4 + \frac{1}{4}b_2\;\left((s_{xx}-s_{yy})^2 +4s_{xy}^2\right)^2 + \frac{1}{4}b_3\;\left(s_{xz}^2+s_{yz}^2\right)^2 \nonumber \\ &+&
    \frac{1}{4}b_4\;s_{zz}^2\;\left((s_{xx}-s_{yy})^2 +4s_{xy}^2\right) +
    \frac{1}{4}b_5\;s_{zz}^2\;\left(s_{xz}^2+s_{yz}^2\right) \\ &+&
    \frac{1}{4}b_6\;\left((s_{xx}-s_{yy})^2 +4s_{xy}^2\right)\;\left(s_{xz}^2+s_{yz}^2\right) \nonumber
\end{eqnarray}
\end{widetext}

using the approximation $\sin\theta \propto \theta$ in (\ref{inv}), the free energy density can then be developed in a power series in $\theta$, without any dependence on the azimuth $\varphi$. This is not surprising as $\theta$ and $\varphi$ can be considered as the modulus and the phase of the Sm-C complex order parameter $\theta \exp(i \varphi)$ \cite{PGDG}~:

\begin{equation}\label{MF}
    F_1 = \frac{1}{2}a_3 \theta^2\ +\ \frac{1}{4}b \theta^4 + ...
\end{equation}

where $a_3 = \alpha_3(T-T_c)$ governs the Sm-A to Sm-C phase transition. This $ \theta^2$ term comes from the 2D-uniaxial invariant while the $\theta^4$ one is the sum of $a_1, a_2, \Omega_1, \Omega_3$ and $b_3$ contributions.

In the mean-field approximation, the  angle $\theta$ behaves like $\frac{b}{\alpha_3}\sqrt{T_c-T}$ in the tilted smectic phases. This looks like previous theories \cite{Z16,O17,L22} which make use only of the 2D-uniaxial invariant missing somewhat the other terms which are at the origin of the Hamaneh-Taylor (H\&T) theory \cite{HT1,HT2,nous}.

From now on we have considered only the case of constant azimuthal angle $\varphi$ like in achiral smectics. As we are dealing with chiral compounds, we can introduce the gradients of $\varphi$ in the $z$ direction to take into account the helicity of chiral smectics.

\subsection{helicity of the first tilted phase}

When the tilt appears at the transition, the first phase that condenses can be the Sm-C$^*$ or the Sm-C$^*_\alpha$, they are distinguished by the value of the pitch of the helix. We have to add to the $F_1$ term the energy contributions of $\overrightarrow{\nabla}\varphi$ coming from the gradients of the oop~:
\begin{widetext}
\begin{equation}\label{nabla}
    F_2 = F_1 - \lambda_{ijklm}\;s_{ij}\nabla_k s_{lm} + \frac{1}{2}k_{ijklmn}\nabla_i s_{jk} \nabla_l s_{mn}
\end{equation}
\end{widetext}
The only gradients which do not perturb the symmetry of the layers are the twisting ones which induce a rotation of the 2D-uniaxial and biaxial invariants around the layer normal. So we assume that the first term linear in $\overrightarrow{\nabla}\varphi$ measures the twisting power of the phase while the second one, quadratic, reflects the cost in twist elastic energy. One finally gets~:
\begin{widetext}
\begin{equation}\label{tw1}
    \lambda_{ijklm}\;s_{ij}\nabla_k s_{lm} = \lambda_1 \left((s_{xx}-s_{yy})\partial_z s_{xy} - s_{xy} \partial_z (s_{xx}-s_{yy}) \right) + \lambda_2 (s_{xz}\partial_z s_{yz} - s_{yz}\partial_z s_{xz})
\end{equation}

\begin{equation}\label{tw2}
    k_{ijklmn}\nabla_i s_{jk} \nabla_l s_{mn} = k_1 \left( (\partial_z s_{xx} - \partial_z s_{yy})^2 + 4\;(\partial_z s_{xy})^2 \right) + k_2 \left( (\partial_z s_{xz})^2 + (\partial_z s_{yz})^2 \right)
\end{equation}
\end{widetext}
when reduced to functions of $\theta$ and $\varphi$ it remains~:
\begin{equation}\label{helix}
    F_2 = F_1 - \lambda \partial_z \varphi +\frac{1}{2} k (\partial_z \varphi)^2
\end{equation}
with the trivial solution $\varphi \;=\; \varphi_0 + q z$ describing an helicoidal rotation at the wave vector $q = \lambda/k$.

The introduction of helicity renormalizes slightly the quadratic term in the free energy leading to a small displacement of the transition temperature $T_c$ to $T_{c1}$ \cite{Martinot}.

Two cases must be distinguished now, depending on the magnitude of the helical pitch $p = 2 \pi/q$.

\subsubsection{small pitch : the Sm-A to Sm-C$^*_\alpha$ phase transition}
When the helical pitch takes values typically from 2 to 8 layers, the Sm-C$^*_\alpha$ phase appears at the transition \cite{tinh}. Although it has been reported in a few compounds a Sm-C$^*_\alpha$ phase with 15 to 50 layers \cite{tinh,HT1}, some caution must exercised as it is on the basis of D.S.C. data taken at $3^\circ \,C \,mn^{-1}$, with a small bump appearing above a large peak, that the phase has been reported \cite{Faye}. To be sure of the phase identification this bump should have been followed at lower speeds as was done in MHPOBC \cite{AED}.

The Sm-C$^*_\alpha$ phase is fundamental in order to get the full sequence of commensurate subphases. It governs the number of layers of the unit cell by means of the azimuth increment $\alpha$ as developed in H\&T theory. One of the experimental challenges risen by this theory is the measurement of $\alpha = 2\pi\,d/p$ by means of the pitch $p$ of the Sm-C$^*_\alpha$ phase. Let us point out that $q = 2 \pi/p$ is the ratio of two polynomials in $\theta$ without any critical dependence~:

\begin{equation}\label{qu}
    q = \frac{\lambda_2 \theta^2 + \lambda_1 \theta^4}{k_2 \theta^2 + 2 k_1 \theta^4}
\end{equation}
we expect it to vary smoothly from $\lambda_2 / k_2$ at the transition to finite values later on. In all the known experiments to date \cite{4,5,Mach,CC,tinh,ellip,Isaert} there is a general trend of increasing $\alpha = q d \simeq q d_0 (1 - \theta^2/2)$ when cooling down at the exception of one compound \cite{Faye,tinh} where the Sm-C$^*_\alpha$ denomination subject to caution. Let us remark that the thermal variation of $q d$ is the product of two terms and cannot be predicted for sure. Close to the transition, the decrease of $d$ when cooling down may be dominant but it seems that later on $q$ will increase and at the end will determine the response.
\subsubsection{large pitch : the direct Sm-A to Sm-C$^*$ phase transition}

When the preferred pitch is typically larger than 0.3 $\mu m$, the tilted phase can be considered locally as a Sm-C$^*$ which precesses slowly around the layer normal. One has then to take into account the macroscopic polarization \cite{Meyer} which rotates too. In order to be coherent, $\vec P$ is defined over a few layers ($\sim 10$), while $s_{jk}$ is relative to one layer and the layer polarization is not a macroscopic quantity \cite{banana}. One has to introduce the macroscopic OOP $Q_{jk} = <s_{jk}>$ which is an average over the same area.  This ensures that when the pitch is smaller than 10 layers, i.e. in the Sm-C$^*_\alpha$ phase, both $\vec P$ and the extra parts in $<s_{jk}>$ vanish. As already stated earlier \cite{JRL,LD,flexo}, the polarization can be formally introduced within linear couplings with the 2D-uniaxial invariant of the OOP, one describing the ferroelectricity \cite{Meyer} and the other the flexoelectricity \cite{OldMeyer,flexo}~:
\begin{eqnarray}\label{P}
    \Delta F_P &=& C_{ijk}\;P_i\;Q_{jk}  + f_{ijkl}\;P_i\;\partial_j\;Q_{kl} \\ \nonumber \\
    &=&  - C \left( P_x\;Q_{yz} - P_y\;Q_{xz} \right)\\ \nonumber \\ \nonumber
    &~& + f \left(P_x\;\partial_z\;Q_{xz} + P_y\;\partial_z\;Q_{yz} \right)
\end{eqnarray}

The total polarization is obtained by the minimization with respect to $P$ of the following energy~:
\begin{equation}
  F = \frac{P_x^2+P_y^2} {2\varepsilon_0\chi} +  \Delta F_P
  \end{equation}

One gets~:
\begin{eqnarray}\label{P}
    \overrightarrow{P} &=& \overrightarrow{P_F} + \overrightarrow{P_f} \\ \nonumber
    P_x &=& \varepsilon_0\chi \left( C\;Q_{yz} - f\;\partial_z Q_{xz}\right) \\ \nonumber
    P_y &=& \varepsilon_0\chi \left( C\;Q_{xz} - f\;\partial_z Q_{yz}\right)
\end{eqnarray}

As expected by symmetry the polarization is an in-plane vector which maximizes $\Delta F_P$ when it is normal to the projection of the director $(Q_{xz}, Q_{yz})$ or equivalently parallel to the gradient $(\partial_z Q_{xz}, \partial_z Q_{yz})$. Both contributions to $\vec P$ are collinear, they follow the helical precession of the director and they change sign with the chirality. The polarization which is measured usually in unwound samples is the ferroelectric one, as the other contribution disappears \cite{Martinot}.

Here again the introduction of the macroscopic ferroelectric polarization renormalizes slightly the quadratic term in the free energy leading to another small displacement of the transition temperature $T_{c1}$ to $T_{c2}$ \cite{LD}. Conversely, the flexoelelectric polarization changes the twist elastic constant and the helical pitch.

\section{The commensurate subphases}

The Sm-C$^*$ and Sm-C$^*_\alpha$ phases are not the only ones encountered in these compounds, when further cooling down a sequence of commensurate phases with unit cells of 1 to 6 layers have been reported \cite{japs,CC2010} which are best described by the distorted clock model \cite{Mach}. The fundamental idea in H\&T theory is that the basic tilted phase obtained below the Sm-A phase is the Sm-C$^*_\alpha$ one, with a short pitch varying from about 2 to 8 layers. When the value of the pitch is close to an integer number of layers, there can be a lock-in of the structure at this integer number at the expense of the twist energy, provided that there is a gain in electrostatic or elastic energy \cite{HT1,HT2,nous}.

Let us enforce the fact that in these subphases, the tilt angle $\theta$ and the $\alpha$ parameter are functions of the temperature only given by the resolution of equations (\ref{MF}) and (\ref{helix}) respectively.
\begin{figure}[!h]
 \begin{center}
{\includegraphics[width=8cm]{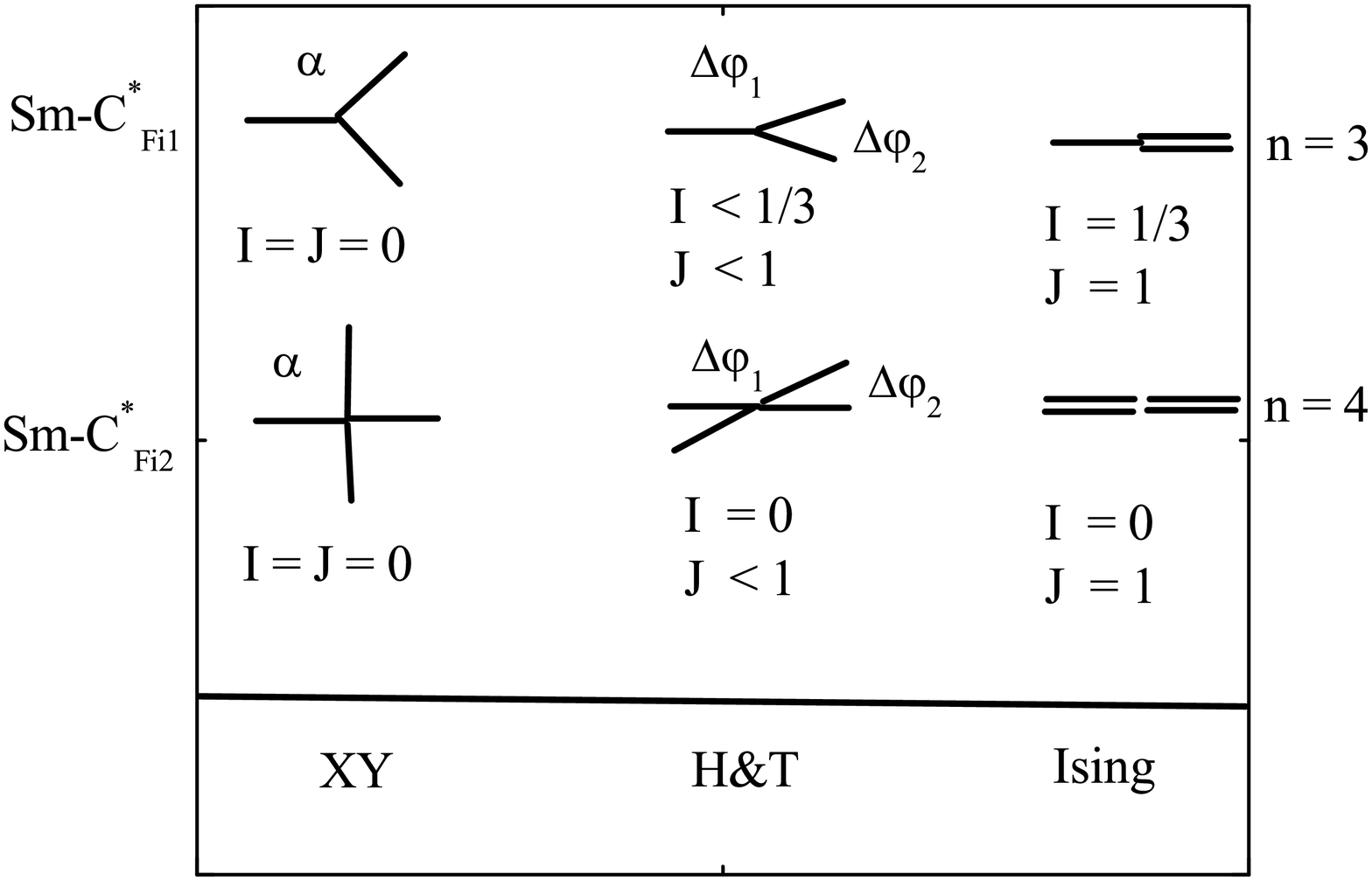}}
 \end{center}
   \caption{Sketch of azimuthal angles in the unit cells of Sm-C$^*_{Fi1}$ (top) and Sm-C$^*_{Fi2}$ phases (bottom). The real distribution in the center is a combination of XY (left) and Ising models (right).} \label{H&T}
\end{figure}

\subsection{Hamaneh-Taylor theory (H\&T)}
The commensurate subphases of the distorted clock model are a compromise between the pure XY clock model \cite{Mach} that requires a regular increase of the azimuth and the old Ising model \cite{Fucku12,Fucku14} that obliges the tilt to lie in one direction of the plane (see e.g. figure (\ref{H&T})). The azimuthal angles in a subphase deviate from the regular XY behavior at the expense of the twist elastic energy while the Ising model favours in-plane anisotropy. H\&T have introduced the angles $\Delta\varphi$ and $\alpha$ to describe the distance from the XY model and the parameter $J =\  \left<\cos 2\varphi\right>\  \leq 1$ to describe the biaxial Ising-like tendency \cite{HT1,HT2}. We have further defined the parameter  $ -1\  \leq\ I = \ \left< \cos \varphi\right>\  \leq 1$ that measures the in-plane uniaxial character \cite{nous}.

 $I$ and $J$ have already been introduced in the definition of the OOP $Q_{ij} = <s_{ij}>$ in equation (\ref{OOP1}), which is a function of the tilt angle $\theta$ and of $\Phi_0$ defined as the angle between the origin of azimuthal angles in the unit cell and the x axis \cite{nous}. The resulting order parameter $Q_{ij}$ is unique ; it is only its expression in a given frame which depends on $\Phi_0$.

\subsection{cost of lock-in : the short range term}
From equation (\ref{helix}) one knows the amount of twist energy per unit volume lost when the wave vector is slightly different from its preferred value $q = 2\pi/p$. It amounts to $\Delta F_t = k(\partial\varphi/\partial z - q)^2/2$. Introducing the azimuthal angle increment $\Delta\varphi$ and the layer thickness $d$, one gets $\Delta F_t = k(\Delta\varphi - qd)^2/2d^2$, $\Delta F_t$ has a zero minimum value when $\Delta\varphi = qd = \alpha$.

This expression averaged over the unit cell of any distorted clock model structure can be related to the short range term $F_{SR}$ introduced empirically by H\&T \cite{HT1}. $F_{SR} = F_0\, <\cos (\Delta\varphi - \alpha)> $  is a positive energy which takes its maximum value $F_0$ in the Sm-C$^*_\alpha$ phase when $\Delta\varphi = \alpha$. On taking $F_0 = k/d^2$ and $\cos(\Delta\varphi-\alpha) \simeq 1-(\Delta\varphi-\alpha)^2/2$ one gets $F_{SR} = F_0 - \Delta F_t$.

We then know how much it costs to lock-in the average increment $<\Delta\varphi> = 2\pi/n$ at a value close to $\alpha$. In the initial clock model this increment was supposed to be constant in the locked-in phases but they would only be commensurate Sm-C$^*_\alpha$ phases without any change in energy. In that case the unit cells are uniaxial without in-plane anisotropy so that the extra parameters $I$ and $J$ are identically null.

\subsection{gain from quadrupolar and dipolar ordering : the $I^2$ and $J^2$ terms}

Uniaxial nematic and Sm-A phases are well known to be ferro-quadrupolar \cite{flexo} i.e. the molecules are arranged at rest so that their microscopic electric dipoles and quadrupoles sum up cooperatively to give a macroscopic quaqrupole density proportional to the uniaxial OOP of the phase~:
\begin{equation}\label{SmA}
    \Theta^0_{ij} = \Theta_a\ S_{ij}
\end{equation}

This can be demonstrated the following way : the quadrupolar density has a quadratic self-energy $\Theta_{ij}^2/2\chi_\Theta$ and is linearly coupled to the OOP by the term $- \Theta_a \Theta_{ij} S_{ij} / 2 \chi_\Theta$, this leads to the result given in equation (\ref{SmA}) and to the expression of the energy gained by the creation of the quadrupolar density~:
\begin{equation}\label{FsmA}
    \Delta F^A_\Theta = - \frac{\Theta^2_a}{3 \chi_\Theta}
\end{equation}

In the Sm-C$^*_\alpha$ phase, this energy decreases with $\theta$ as~:
\begin{equation}\label{Fsma}
    \Delta F^\alpha_\Theta = - \frac{\Theta^2_a}{3 \chi_\Theta} \ \left(1 - \frac{3}{2} \sin^2 \theta\right)^2
\end{equation}
the lock-in to commensurate subphases allows to recover a part of this loss. In the uniaxial approximation we are using, the quadrupolar energy in the Sm-C$^*$ phase is the same as in the Sm-A, $\Delta F^C_\Theta = \Delta F^A_\Theta$ so~:
\begin{widetext}
\begin{equation}\label{FsmC}
    \Delta F^C_\Theta = - \frac{\Theta^2_a}{3 \chi_\Theta} \ \left( \left(1 - \frac{3}{2} \sin^2 \theta \right)^2 + \frac{3}{4} \sin^4 \theta + 3 \sin^2 \theta \cos^2 \theta \right)
\end{equation}
\end{widetext}
a straightforward generalization to any value of $I$ and $J$ reads~:
\begin{widetext}
\begin{equation}
    \Delta F^{IJ}_\Theta = - \frac{\Theta^2_a}{3 \chi_\Theta} \ \left( \left(1 - \frac{3}{2} \sin^2 \theta \right)^2 + \frac{3\ J^2}{4} \sin^4 \theta + 3\ I^2\;\sin^2 \theta \cos^2 \theta \right)
\end{equation}
\end{widetext}
We are eventually left with the gain of lock-in which reads~:
\begin{widetext}
\begin{eqnarray}\label{ouf}
\nonumber
    \Delta F = \Delta F^{IJ}_\Theta - \Delta F^\alpha_\Theta &=& - \frac{\Theta^2_a}{ \chi_\Theta} \ \left( \frac{J^2}{4} \sin^4 \theta + I^2\,\sin^2 \theta \cos^2 \theta \right) \\
    &=&\ - F_0\ (\eta\,J^2 + \gamma_1\,\sqrt{\eta}\,I^2) \approx \ - F_0\ ( \widetilde{\eta} \,\theta^4\,J^2 + \widetilde{\gamma_1}\,\theta^2\,I^2)
\end{eqnarray}
\end{widetext}
When completed by the similar term due to the presence of the macroscopic polarization $\overrightarrow{P_S}$ if $I \neq 0$ as developed by Dhaouadi et al. \cite{nous}, one gets~:

\begin{eqnarray}
\label{eq3}
P_S  &=&  \varepsilon_0  \chi  C I \theta \nonumber \\
\\
\Delta \tilde {F}_P   &=&  - \frac{P_S^{ 2}} {2\varepsilon_0\chi} =  - \frac{\varepsilon _0  \chi  C^2 \theta
^2 }{2}I^2 \nonumber
\end{eqnarray}
the full energy gain reads with this n
ew $I^2$ term~:
\begin{widetext}
\begin{equation}\label{ouf}
    \Delta \widetilde{F} =\ - F_0\ (\eta\,J^2 + \gamma\,\sqrt{\eta}\,I^2)
    \approx \ - F_0\ ( \widetilde{\eta} \,\theta^4\,J^2 + \widetilde{\gamma}\,\theta^2\,I^2)
\end{equation}
\end{widetext}
\subsection{balance between short range loss and long range gain}

When a phase described by the distorted clock model appears, it is characterized by non zero values of $I$ and $J$ that minimize the following energy at a negative value~:
\begin{equation}\label{HT0}
    F = F_0 \left [ \frac{1}{2} \left < \Delta\varphi - \alpha\right >^2  - \eta J^2 - \gamma \sqrt{\eta} I^2\right]
\end{equation}

The phase diagrams in the plane ($0 \leqslant \alpha \leqslant \pi$ , $0 \leqslant \eta \leqslant 1$) have been computed by H\&T \cite{HT1,HT2} with the $J^2$ term and Dhaouadi \cite{nous} ($I^2$) together with their behavior under an applied electric field (-$I.E$) in the last case.

Let us remark that up to now we have discussed the commensurate subphases in the unwound geometry although we know that they are all precessing around the layer normal. We propose to treat this problem now and compare our results with the well known pitch inversion at the Sm-C$^*_{Fi1}$ to Sm-C$^*_{Fi2}$ phase transition.

\begin{figure}[!h]
  \includegraphics[width=8cm]{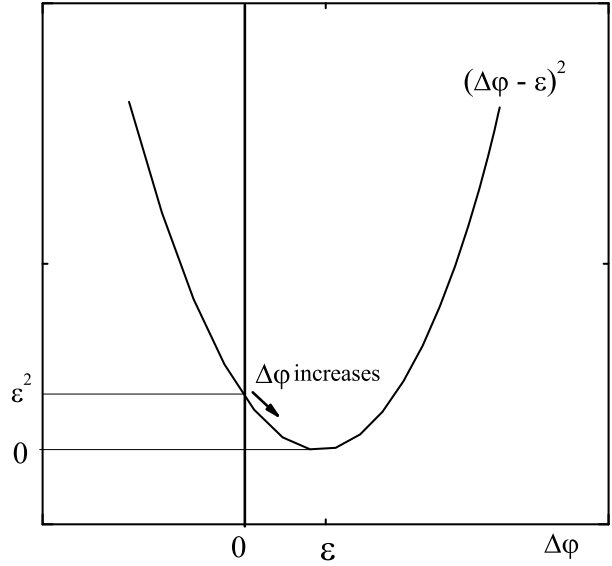}\\
  \caption{Sketch of the short range term expressing the cost of lock-in from Sm-C$^*_\alpha$ ($\Delta\varphi = \varepsilon$) to Sm-C$^*$ phase ($\Delta\varphi = 0$). A small positive increment $\delta(\Delta\varphi)$ would reduce the cost.}\label{HC}
\end{figure}
\subsection{helicity of the subphases}

\begin{figure}[!h]
  \includegraphics[width=8cm]{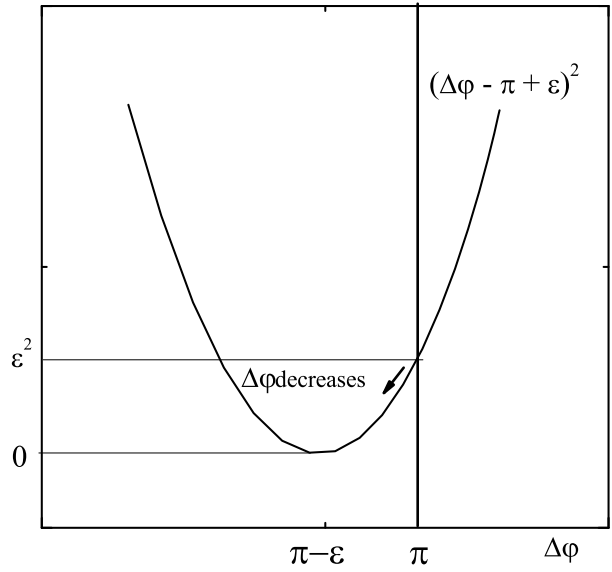}\\
  \caption{Sketch of the short range term expressing the cost of lock-in from Sm-C$^*_\alpha$ ($\Delta\varphi = \pi - \varepsilon$) to Sm-C$^*_A$ phase ($\Delta\varphi = \pi$). A smaller $\Delta\varphi$ would reduce the cost.}\label{HCA}
\end{figure}

We have considered when computing the phase diagrams that the unit cells of the commensurate subphases were frozen and did not rotate while the subjacent Sm-C$^*_\alpha$ phase experiences a strong spontaneous twist. We first examine graphically the simple cases of the Sm-C$^*$ and Sm-C$^*_A$ phases where the azimuth increase between layers is equal to $\Delta \varphi = 0$ or $\Delta \varphi = \pi$ while the Sm-C$^*_\alpha$ phase that would take place otherwise has a small value $\alpha = \varepsilon$ in the first case (figure \ref{HC}) or is close to $\pi$ in the second ($\alpha = \pi - \varepsilon$ figure \ref{HCA}). The figures show that the cost in twist energy of the lock-in will be reduced if the Sm-C$^*$ rotates in the same direction as the
Sm-C$^*_\alpha$ while the Sm-C$^*_A$ has to take the opposite sense. This illustrates the general trend observed in the experiments \cite{8} that the sense of the helix is opposite in the two phases for a given compound. We have thus shown that the would-be Sm-C$^*_\alpha$ phase exercises a kind of torque on the azimuthal angle $\Phi_0$ with a non trivial sign. It has to be completed by a spontaneous twist we will develop now.

We first take into account the preceding remarks by replacing $\Delta F_t = F_0\,(\Delta\varphi - \alpha)^2/2$ by $\widetilde{\Delta F_t} = F_0\,(\Delta\varphi + d\ \partial_z \Phi_0 - \alpha)^2/2$. We then state that the macroscopic angle $\Phi_0$ is subject to the same laws than the microscopic $\varphi$~:
\begin{widetext}
\begin{equation}\label{helix2}
    \widetilde{F_t} = F_0\,(\Delta\varphi + d\ \partial_z \Phi_0 - \alpha)^2/2 - \Lambda\ \partial_z \Phi_0 +\frac{1}{2} K (\partial_z \Phi_0)^2
\end{equation}
\end{widetext}
replacing $F_0$ by $k/d^2$, one has to minimize~:
\begin{equation}\label{H}
    \frac{1}{2}(k+K)\ (\partial_z\,\Phi_0)^2 + \left( \frac{k}{d}\,\left(\Delta\varphi - \alpha\right) - \Lambda \right) \partial_z\,\Phi_0
\end{equation}
yielding~:
\begin{equation}
    \partial_z\,\Phi_0 = \frac{\Lambda -k(\Delta\varphi - \alpha)/d}{k+K}
\end{equation}
thus the wave vector of the macroscopic helicity is given as usual by the ratio of the twist sources to the rigidity. We already know that the pitches are much larger than in the Sm-C$^*_\alpha$ phase so one may assume safely $K \gg k$. The source in the numerator is made of two terms. The first one is the intrinsic twisting power $\Lambda$ which we take as having the same sign as the microscopic one $\lambda$, this remaining to be checked in real world. The second one corresponds to the discussion we just developed with the help of figures (\ref{HC}) and (\ref{HCA}).

Let us briefly determine the sign of $\partial_z \Phi_0$ in the different commensurate subphases~:
\begin{enumerate}
  \item in Sm-C$^*$ the second term dominates and has the sign of $\alpha$.
  \item in Sm-C$^*_A$ the second term dominates with the opposite sign.
  \item in Sm-C$^*_{Fi1}$ if the angle $\mu$ is much larger than $2 \pi/3$ as reported in literature \cite{nous}, the opposite sign comes again.
  \item the Sm-C$^*_{Fi2}$ is more involved as it seems that the first source dominates and in our hypothesis on the sign of $\Lambda$, the helix has the same sign as $\alpha$.
\end{enumerate}
So we have found that usual sequence with a sign change in the middle of the range is well explained in our theory.

\section{ Conclusion}

We have demonstrated that the sequence of distorted commensurate phases observed in tilted chiral smectics is explained by the gain in electrostatic energy due to the lock-in of the unit cell to a number of layers which is the integer closest to the ratio pitch over thickness of the subjacent Sm-C$^*_\alpha$ phase. We also explain the sign change of the helicity in the middle of the sequence by a balance between two twist sources.

\appendix
\section{$T_c$ shifts}

\subsection{due to the helicity}

remembering equation (\ref{helix}) and introducing the preferred wave vector $q = \lambda/k$, one gets $\widetilde{F_2} = F_1 - kq^2/2 = F_1 - \lambda^2/2 k$. $F_1$ is an even polynomial in $\theta$ as well as the correction term, the leading term reads now after equation (\ref{MF}) $\alpha_3(T-T_c)\theta^2/2 - \lambda_2^2/2 k_2 = \alpha_3(T-T^*_c)\theta^2/2 $ with a straightforward definition of $T^*_c > T_c$.

\subsection{due to the polarization}

Here again the correction term due to the macroscopic ferroelectric polarization can be expressed as $\Delta F^1_P = - \chi C^2 \theta^2$ leading to another shift in the transition temperature $\Delta T_c^* = \chi C^2 / \alpha_3$.

\section{biaxiality vs uniaxiality}

All the unwound phases in the distorted clock model are biaxial with the in-plane C2 axis as one eigenaxis. Their OOP $S_{ij}$ and the rotation matrices should read~:
\begin{equation}
 S_{ij}   = \left( \begin{array}{c c c}
 -a  & 0  & 0  \\
 0  & -b  & 0  \\
 0  & 0 & a+b
\end{array}  \right)
\ P^\theta_{ij}   = \left( \begin{array}{c c c}
 \cos \theta  & 0  & \sin \theta  \\
 0  & 1  & 0  \\
 -\sin \theta   & 0 & \cos \theta
\end{array}  \right)
\ P^\varphi_{ij}   = \left( \begin{array}{c c c}
 \cos \varphi   & \sin \varphi   & 0  \\
 -\sin \varphi  & \cos \varphi  & 0  \\
 0  & 0 & 1
\end{array}  \right)
\end{equation}
when one executes a rotation of angle $\theta$ around this axis and another of angle $\varphi$ around $3$, one gets~ :
\begin{widetext}
 \begin{eqnarray}
\nonumber  Q_{ij} &=& \frac{3}{2} \left( a \cos 2\theta + b \cos^2\theta \right) \left( \begin{array}{c c c}
 -1/3  & 0  & 0  \\
 0  & -1/3  & 0  \\
 0  & 0 & +2/3
\end{array} \right)\\
 &+& \frac{1}{2}\
(- a \cos 2\theta + b (1 + \sin^2\theta) )
\left( \begin{array}{c c c}
 \cos 2\varphi  & \sin 2\varphi  & \ 0\   \\
 \sin 2\varphi  & -\cos 2\varphi  & \ 0\   \\
 0  & 0 &\  0  \end{array} \right)
 \\ \nonumber
 ~ &-& (2 a + b) \sin \theta \cos \theta
 \left( \begin{array}{c c c}
  0 & 0  & \ \cos \varphi\   \\
 0  & 0  & \ \sin \varphi\   \\
 \cos \varphi  & \sin \varphi &\  0\
\end{array} \right)
\end{eqnarray}
\end{widetext}

this differs only slightly in the coefficients from the uniaxial form we have used. We took  $a = b = 1/3$, we could have been closer to reality with $a = 1/3, \ b = 1/3 + \delta$ because of $\varphi$ fluctuations evidenced by conoscopy under field \cite{japs}.

\begin{acknowledgements}
We wish to acknowledge the support of CRPP's \emph{pot commun} and the help of the Tunis group for this work.
\end{acknowledgements}

\newpage

\bibliography{qunu}

\end{document}